\documentclass[conference]{IEEEtran}
\IEEEoverridecommandlockouts
\usepackage{cite}
\usepackage{amsmath,amssymb,amsfonts}
\usepackage{algorithmic}
\usepackage{graphicx}
\usepackage{textcomp}
\usepackage{xcolor}
\usepackage{flushend}
\def\BibTeX{{\rm B\kern-.05em{\sc i\kern-.025em b}\kern-.08em
    T\kern-.1667em\lower.7ex\hbox{E}\kern-.125emX}}

\newtheorem{lemma}{Lemma}
\usepackage{url}

\allowdisplaybreaks
\begin{document}

\title{\huge 
Energy Efficiency Optimization of  Finite Block Length STAR-RIS-aided MU-MIMO Broadcast Channels 
}

\author{
   \IEEEauthorblockN{Mohammad Soleymani$^1$, Ignacio Santamaria$^2$, Eduard Jorswieck$^3$, Robert Schober$^4$, Lajos Hanzo$^5$}\vspace{.2cm}
   \IEEEauthorblockA{
   $^1$University of Paderborn, Germany,  $^2$Universidad de Cantabria, Spain,
   $^3$ Technische Universit\"at Braunschweig, Germany\\
   $^4$Friedrich Alexander University of Erlangen-Nuremberg, Germany,
   $^5$University of Southampton, United Kingdom\vspace{.2cm}
   \\
                     Email: \small{\protect\url{mohammad.soleymani@uni-paderborn.de}}, \small{\protect\url{i.santamaria@unican.es}}, \small{\protect\url{jorswieck@ifn.ing.tu-bs.de}}\\
                     \small{\protect\url{robert.schober@fau.de}}, \small{\protect\url{lh@ecs.soton.ac.uk}}
}
}
\maketitle

\begin{abstract}
Energy-efficient designs are proposed for multi-user (MU) multiple-input multiple-output (MIMO) broadcast channels (BC), assisted by simultaneously transmitting and reflecting (STAR) reconfigurable intelligent surfaces (RIS) operating at finite block length  (FBL). In particular, we maximize the sum energy efficiency (EE), showing that STAR-RIS can substantially enhance it. Our findings demonstrate that the gains of employing STAR-RIS increase when the codeword length and the maximum tolerable bit error rate decrease, meaning that a STAR-RIS is more energy efficient in a system with more stringent latency and reliability requirements. 
\end{abstract}


\section{Introduction}
Enhancing energy efficiency (EE) has always been among the primary goals of wireless communication systems, aming for realizing green communications and reducing operational expenses \cite{wang2023road, gong2022holographic}. Additionally, the next generations of wireless communication systems should be able to reduce latency while simultaneously increasing reliability \cite{vaezi2022cellular}. Hence, we aim to propose an energy-efficient design for ultra-reliable low-latency communication (URLLC) systems to address these issues. 

A promising tool to enhance EE, latency, and reliability is constituted by reconfigurable intelligent surfaces (RIS) \cite{wu2021intelligent, di2020smart}. The EE benefits of RIS were investigated in \cite{lyu2023energy, soleymani2024energy, soleymani2022rate, niu2023active, soleymani2022noma, fotock2023energy}, showing that RIS improves the EE of various multi-user (MU) systems like cell-free systems, broadcast channels (BCs), and multiple access channels when considering Shannon's capacity that assumes that the coding block length approaches infinity. However, in the face of low-latency demands, we often have to utilize finite block lengths (FBL), making the classical Shannon rate an inaccurate metric to measure the achievable data rate \cite{polyanskiy2010channel, erseghe2016coding}. The benefits of RIS in systems operating at FBL coding were shown in \cite{soleymani2024rate2, li2021aerial, vu2022intelligent, xie2021user, almekhlafi2021joint, soleymani2023spectral,   pala2023spectral,  soleymani2023optimization, katwe2024rsma, soleymani2024optimization, soleymani2024rate}. More specifically, \cite{soleymani2024rate2} showed that RIS can enlarge the rate region of a multiple-input single-output (MISO) BC. The authors of \cite{xie2021user} used RIS to minimize the latency in a BC. The paper \cite{soleymani2023optimization} proposed algorithms to improve spectral efficiency and EE of multi-cell MISO URLLC BCs. In \cite{katwe2024rsma}, the authors maximized the sum rate of RIS-aided MISO BCs. Finally, the papers \cite{soleymani2024optimization, soleymani2024rate}  demonstrated that RIS enhances the EE and spectral efficiency of a multi-cell MU multiple-input multiple-output (MIMO) BC, supporting multiple data streams per user.

There are various nearly passive RIS architectures, including both reflective RISs and simultaneously transmitting and reflecting (STAR) RIS to support different applications. A reflective RIS assists a communication link only if both transceivers are in the reflection space of the RIS, while STAR-RIS provides an omnidirectional coverage  \cite{zhang2022intelligent, liu2021star}. Hence, a STAR-RIS supports a broader range of applications than a purely reflective RIS. For instance, a STAR-RIS can be fixed to a window, covering both indoor and outdoor users. However, a reflective RIS with the same position can support only indoor (or outdoor users) as it covers only half of the space. In \cite{jorswieck2025urllc}, it was illustrated that STAR-RIS improves the max-min rate of a MISO BC. The ability of STAR-RIS to enhance the SE and EE of multi-cell MISO BCs was shown in \cite{soleymani2023spectral}.

In this paper, we focus on the EE performance of FBL STAR-RIS-assisted MU-MIMO systems. To this end, we develop an energy-efficient algorithm to maximize the sum EE of users in STAR-RIS-aided MU-MIMO URLLC BCs. In \cite{soleymani2024optimization}, we obtained a closed-form expression for the achievable rate of MU-MIMO systems with FBL processing supporting multiple data-streams per user and developed spectral-efficient and energy-efficient schemes for MU-MIMO RIS-assisted systems by solving the minimum rate, sum weighted rate, minimum EE and global EE maximization problems. In this paper, we solve the weighted sum EE maximization problem of STAR-RIS-aided URLLC systems and consider both mode switching (MS) and energy splitting (ES) schemes to operate a STAR-RIS. The weighted sum EE maximization problem is more complex than minimum EE and global EE maximization, since its objective function is a summation of multiple fractional functions. Solving the weighted sum EE maximization, we show that STAR-RIS outperforms a reflective RIS when there are users in both the reflection and refraction spaces of the RIS. Indeed, in this case, a reflective RIS cannot assist all the users, reducing efficiency. The gains provided by STAR-RIS escalate when the codeword length and maximum tolerable detecting error rate decrease, indicating higher benefits for more stringent latency and reliability demands.

The rest of this paper is organized as follows. Section \ref{sec:smps} presents the system model, states the rate and EE expressions, and formulates the optimization problem we tackle in this paper. Section \ref{sec-iii} presents our solution. We provide numerical results in Section \ref{sec:ni} and conclude the paper in Section \ref{sec:conc}.

{\em Notations}: We represent scalars/vectors/matrices by $x/{\bf x}/{\bf X}$. A zero-mean complex Gaussian vector ${\bf x}$ with covariance matrix ${\bf X}$ is denoted by ${\bf x}\sim\mathcal{CN}({\bf 0}, {\bf X} )$. Moreover, $Q^{-1}$ and ${\bf I}$ denote the inverse Q-function for Gaussian signals and an identity matrix, respectively. 
Finally, $\text{Tr}({\bf X})$, $|{\bf X} |$, and $\mathfrak{R}\{x\}$ take, respectively, the trace of ${\bf X} $, the determinant of ${\bf X} $, and the real value of $x$.

\section{System Model and Problem Statement}
\label{sec:smps}
\begin{figure}[t]
	\centering
    \includegraphics[width=.9\linewidth]{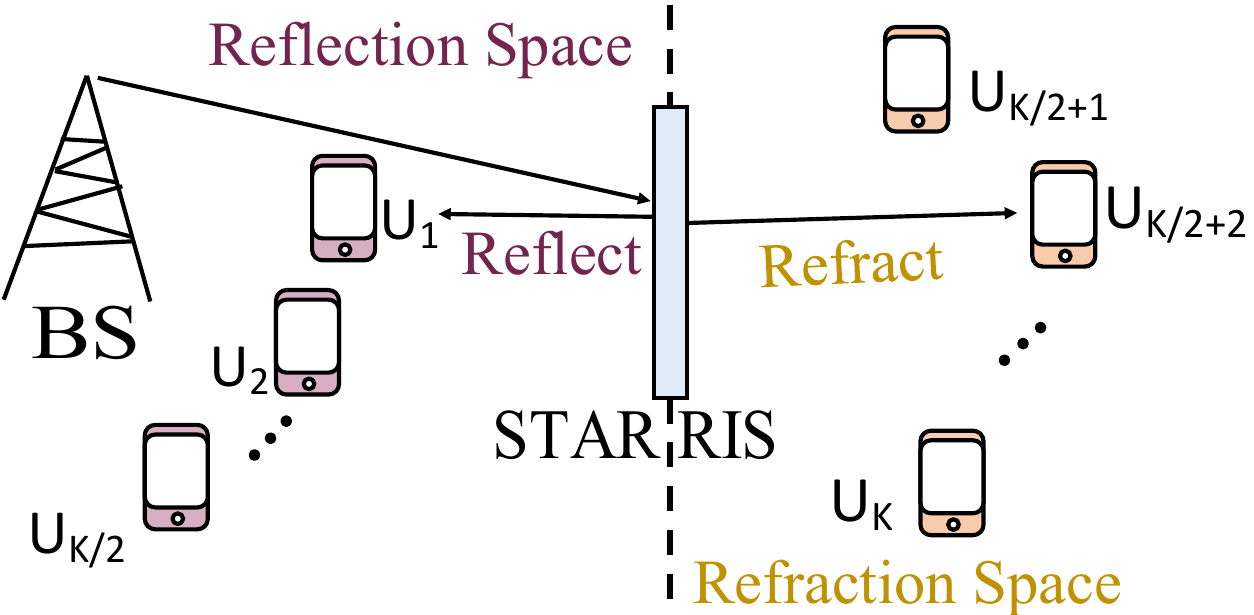}
	\caption{A broadcast channel assisted by STAR-RIS. }
	\label{fig:network-model}
\end{figure}

We consider a MIMO BC, aided by a STAR-RIS, which is depicted in Fig. \ref{fig:network-model}. More specifically, we assume that a multiple-antenna base station (BS) serves $K$ multiple-antenna users, and a STAR-RIS assists their communication. We denote the number of transmit antennas at the BS by $N_{BS}$ and the number of receive antennas at each user by $N_u$. Moreover, we assume that the BS employs FBL coding to transmit multiple data streams to each user. 

\subsection{Channel Model}

The channel between the BS and user $k$ is modeled as  \cite{mu2021simultaneously}
\begin{eqnarray}
	\label{eq:cm}
	{\bf H}_k = {\bf F}_k {\bf \Theta}^{r/t} {\bf G} + {\bf D}_k, 
\end{eqnarray}
where ${\bf D}_k$ is the channel between the BS and user $k$, ${\bf F}_k$ is the channel between user $k$ and the STAR-RIS, and ${\bf G}$ is the channel between the BS and the STAR-RIS. The matrices ${\bf \Theta}^{r}=\text{diag}\left(\theta_1^r,\theta_2^r,\cdots,\theta_M^r\right)$ and ${\bf \Theta}^{t}=\text{diag}\left(\theta_1^t,\theta_2^t,\cdots,\theta_M^t\right)$ contain the RIS coefficients in the reflection and refraction spaces, respectively. 
Note that if a user is in the reflection (or refraction) space, its channel can only be optimized through ${\bf \Theta}^{r}$ (or ${\bf \Theta}^{t}$). We consider a nearly passive RIS, meaning that the amplitude of the RIS coefficients should obey
\begin{equation}\label{eq-12}
    |\theta^r_m|^2+|\theta^t_m|^2\leq 1,\,\,\, \forall m.
\end{equation}
In \cite{liu2022star}, three operational modes were proposed for STAR-RIS, namely energy splitting (ES),  mode switching (MS), and time switching (TS). In the ES mode, each STAR-RIS element operates in both the reflection and refraction spaces, yielding a better performance at the cost of higher operational and computational complexity. On the contrary, in the MS and TS schemes, each element operates only in either the reflection or the refraction space. In particular,  when utilizing the MS scheme in this paper, we divide the STAR-RIS elements into two groups. The elements in the first group ($1\leq m\leq m_g$) operate only in the reflection mode, meaning that $\theta^{t}_m=0$ for all $m\leq m_g$. By contrast, the elements of the other group ($m_g< m\leq M$) operate only in the refraction mode, resulting in $\theta^{r}_m=0$ for all $m> m_g$. Although the MS scheme has lower complexity than the ES scheme, it performs almost similarly in many practical scenarios, as shown in \cite{soleymani2023noma2, soleymani2023energy}. In this paper, we consider only the ES and MS schemes. However, our solutions can be easily extended to the TS scheme. 

\subsection{Signal Model and Problem Statement}
The BS leverages superposition coding to transmit data to users. Thus, the signal transmitted at the BS is  
\begin{eqnarray}
	{\bf x} =  \sum_{k=1}^K {\bf \Gamma}_k {\bf t}_k \label{eq:txsig}, 
\end{eqnarray}
where ${\bf t}_k\sim\mathcal{CN}({\bf 0}, {\bf I} )$ is the private message of user $k$, and ${\bf \Gamma}_k$ is the beamforming matrix corresponding to ${\bf t}_k$.
Therefore, the data received at user $k$ is 
\begin{eqnarray}\label{eq:recsig}
	{\bf y}_k = {\bf H}_k \sum_{j=1}^K {\bf \Gamma}_j {\bf t}_j+ {\bf z}_k, 
\end{eqnarray}
where ${\bf z}_k \sim \mathcal{CN}(0,\sigma^2{\bf I} )$ is the additive noise. 
Each user leverages treating interference as noise (TIN) to detect its message. Hence, the achievable rate of user $k$ is \cite{polyanskiy2010channel, soleymani2024optimization}
\begin{multline}
    	r_{k}\! = \log\!\left| {\bf I}\! +\!\! \left(\!\!\sigma^2{\bf I}\!+\sum_{i\neq k} {\bf H}_k{\bf \Gamma}_i{\bf \Gamma}_i^H{\bf H}_k^H\!\! \right)^{-1}\!\!\!\! {\bf H}_k{\bf \Gamma}_k{\bf \Gamma}_k^H{\bf H}_k^H \right|\\
        \!- \frac{Q^{-1}(\epsilon)
         \sqrt{ 
        {\!2\text{Tr}\!\left
    (\!{\bf H}_k{\bf \Gamma}_k{\bf \Gamma}_k^H{\bf H}_k^H\!
    \left[\!\sigma^2{\bf I}\!+\!
    \sum_{i} {\bf H}_k{\bf \Gamma}_i{\bf \Gamma}_i^H{\bf H}_k^H
    \right]^{-1}\! \right
    )}
    }}{\sqrt{l}}\!, \!\!\!\!\!\!
    \label{eq:cr}
\end{multline}
where  $\epsilon$ is the maximum tolerable detecting error rate, and $l$ is the codeword length.
Furthermore, the EE of user $k$ is 
\begin{equation}
    e_k=\frac{r_k}{P_s+\beta\text{Tr}({\bf \Gamma}_k{\bf \Gamma}_k^H )},
\end{equation}
where $\beta$ is the power efficiency of the transmitter at the BS, and $P_s$ is the static power dissipated by transmitting data to user $k$, given by \cite[Eq. (27)]{soleymani2022improper}.

\subsection{Problem Statement}

We aim at maximizing the sum EE of users, resulting in the following optimization problem 
\begin{subequations}\label{opt}
    \begin{align}    
	\max\limits_{\{{\bf \Gamma}_k\}_{\forall k}, {\bf \Theta}^r, {\bf \Theta}^t} & \sum_{k } \alpha_k e_k  \\
	\label{eq:qosi}\textrm{s.t.}\hspace{.5cm} & r_k\geq \bar{r}_k^{th}\\
    & \sum_{k=1}^K\text{Tr}\left( {\bf \Gamma}_k{\bf \Gamma}_k^H \right)\! \leq\! P \label{eq:con1} \\
	& |\theta^r_m|^2+|\theta^t_m|^2\leq 1,\,\,\, \forall m, \label{eq:con2}
\end{align}
\end{subequations}
where $\alpha_k>0$ is the corresponding weight for user $k$, $P$ is the BS power budget, $\bar{r}_k^{th}$ is the minimum operational rate for user $k$, and the constraint in \eqref{eq:qosi} corresponds to the quality of service for user $k$. The optimization problem in \eqref{opt} is a fractional matrix programming (FMP) problem. To solve \eqref{opt}, we leverage the optimization framework proposed in \cite{soleymani2025framework}.  A detailed solution is provided in Section \ref{sec-iii}.

\section{Proposed Solution}\label{sec-iii}
\label{sec:aora}
This section provides a suboptimal solution for \eqref{opt}, utilizing the framework in \cite{soleymani2025framework} and alternating optimization. To solve \eqref{opt}, we first optimize the beamforming matrices when ${\bf \Theta}^r$ and ${\bf \Theta}^t$ are kept fixed to ${\bf \Theta}^{r^{(n)}}$ and ${\bf \Theta}^{t^{(n)}}$, respectively, where $n$ is the iteration index. We then alternate the optimization variables and update ${\bf \Theta}^r$ and ${\bf \Theta}^t$ while keeping the beamforming matrices fixed at ${\bf \Gamma}^{(n+1)}_k$ for all $k$. 
We solve each optimization problem in a separate subsection below. 

\subsection{Beamforming Optimization}

When ${\bf \Theta}^r$ and ${\bf \Theta}^t$ are fixed, \eqref{opt} is simplified to
    \begin{align} \label{opt-w}   
	\max\limits_{\{{\bf \Gamma}_k\}_{\forall k}} & \sum_{k } \alpha_ke_k &
	\textrm{s.t.}\hspace{.5cm} & \eqref{eq:qosi}, \eqref{eq:con1},
\end{align}
which is a non-convex FMP problem. To solve \eqref{opt-w}, we first compute a concave lower bound for $r_k$, presented in Lemma \ref{lem:1}, and then apply \cite[Lemma 2]{soleymani2025framework} to approximate \eqref{opt-w} with a convex problem.
\begin{lemma}[\!\!\cite{soleymani2024optimization}]
	\label{lem:1}
	For every feasible  ${\bf \Gamma}_k$ for all $k$, the inequality in \eqref{eq-10-low}, shown at the top of the next page, holds, where $a_{k}$, ${\bf A}_{kj}$, and ${\bf B}_{k}$ are, respectively, given by \eqref{eq-a}, \eqref{eq-Akj}, and \eqref{eq-b} on the top of the next page. Moreover, $I\leq\min(N_{BS}, N_u)$ is the maximum number of data streams per user, and $\eta_k$ is given by 
\begin{multline*}
        \eta_k=2\text{Tr}\left
    (\mathbf{H}_{k}^{(n)}{\bf \Gamma}_k^{(n)}{\bf \Gamma}_k^{(n)^H}\mathbf{H}_{k}^{(n)^H}\times
    \right.\\    \left.
    \left[\sigma^2{\bf I}+
    \sum_{i} \mathbf{H}_{k}^{(n)}{\bf \Gamma}_i^{(n)}{\bf \Gamma}_i^{(n)^H}\mathbf{H}_{k}^{(n)^H}
    \right]^{-1} \right
    ),
\end{multline*}
where $\mathbf{H}_{k}^{(n)}$ is given by inserting ${\bf \Theta}^{(n)^r}$ and ${\bf \Theta}^{(n)^t}$ in \eqref{eq:cm}.
\end{lemma}
\begin{figure*}[t]
    \begin{align}
    \label{eq-10-low}
    r_{k}& \geq \tilde{r}_{k}= a_{k}
+2\sum_{j}\mathfrak{R}\left\{\text{{Tr}}\left(
{\bf A}_{kj}{\bf \Gamma}_{j}^H
\mathbf{H}_{k}^{(n)^H}\right)\right\}
-
\text{{Tr}}\left(
{\bf B}_{k}\left(\sigma^2{\bf I}
+\sum_j\mathbf{H}_{k}^{(n)}{\bf \Gamma}_{j}{\bf \Gamma}_{j}^H\mathbf{H}_{k}^{(n)^H}\right)
\right)
\\
a_{k}&=\ln\left|{\bf I}+\left(\!\sigma^2{\bf I}+\sum_{j\neq k}
\mathbf{H}_{k}^{(n)}{\bf \Gamma}_{j}^{(n)}{\bf \Gamma}_{j}^{(n)^{H}}\mathbf{H}_{k}^{(n)^H}\right)^{-1}\mathbf{H}_{k}^{(n)}{\bf \Gamma}_{k}^{(n)}{\bf \Gamma}_{k}^{(n)^{H}}\mathbf{H}_{k}^{(n)^H}\right|
\nonumber\\
&\hspace{1.5cm}-\text{{Tr}}\left(\left(\sigma^2{\bf I}+\sum_{j\neq k} \mathbf{H}_{k}^{(n)}{\bf \Gamma}_{j}^{(n)}{\bf \Gamma}_{j}^{(n)^{H}}\mathbf{H}_{k}^{(n)^H}\right)^{-1}\mathbf{H}_{k}^{(n)}{\bf \Gamma}_{k}^{(n)}{\bf \Gamma}_{k}^{(n)^{H}}\mathbf{H}_{k}^{(n)^H}
\right)
-\frac{Q^{-1}(\epsilon)({\eta_k}+2I 
)}{2\sqrt{l\eta_k}},
\label{eq-a}
\\
\label{eq-Akj}
{\bf A}_{kj}&=\left\{\begin{array}{ll}
\left(\sigma^2{\bf I}+\sum_{j\neq k}\mathbf{H}_{k}^{(n)}{\bf \Gamma}_{j}^{(n)}{\bf \Gamma}_{j}^{(n)^{H}}\mathbf{H}_{k}^{(n)^H}\right)^{-1}{\mathbf{H}}_{k}^{(n)} {\bf \Gamma}_{k}^{(n)}
&
\text{if}
\,\,i=l,
\,j=k\\
\frac{Q^{-1}(\epsilon)}{\sqrt{l\eta_k}}
( \sigma^2{\bf I}+\sum_{j}\mathbf{H}_{k}^{(n)}{\bf \Gamma}_{j}^{(n)}{\bf \Gamma}_{j}^{(n)^{H}}\mathbf{H}_{k}^{(n)^H})^{-1}\mathbf{H}_{k}^{(n)}{\bf \Gamma}_{j}^{(n)}
&
\text{otherwise}
\end{array}\right.
\\
{\bf B}_{k}&\!=\!
\frac{Q^{-1}(\epsilon)}{\sqrt{l\eta_k}}\!\!
\left(\!\!
\sigma^2{\bf I}\!+\!\!\sum_{j}\mathbf{H}_{k}^{(n)}{\bf \Gamma}_{j}^{(n)}{\bf \Gamma}_{j}^{(n)^{H}}\mathbf{H}_{k}^{(n)^H}\!\!\!\right)^{-1}
\!\!\!\!
\left(\!\!\sigma^2{\bf I}\!+\!\!\sum_{j\neq k}\mathbf{H}_{k}^{(n)}{\bf \Gamma}_{j}^{(n)}{\bf \Gamma}_{j}^{(n)^{H}}\mathbf{H}_{k}^{(n)^H}\!\!\!\right)
\!\!\!
\left(\!\!\sigma^2{\bf I}\!+\!\!\sum_{j}\mathbf{H}_{k}^{(n)}{\bf \Gamma}_{j}^{(n)}{\bf \Gamma}_{j}^{(n)^{H}}\mathbf{H}_{k}^{(n)^H}\!\!\right)^{-1}\!
\nonumber
\\
&\hspace{.6cm}
+\left(\sigma^2{\bf I}\!+\!\sum_{j\neq k}\mathbf{H}_{k}^{(n)}{\bf \Gamma}_{j}^{(n)}{\bf \Gamma}_{j}^{(n)^{H}}\mathbf{H}_{k}^{(n)^H}\right)^{-1}-\left(\sigma^2{\bf I}+\sum_{j}\mathbf{H}_{k}^{(n)}{\bf \Gamma}_{j}^{(n)}{\bf \Gamma}_{j}^{(n)^{H}}\mathbf{H}_{k}^{(n)^H}\right)^{-1}
\label{eq-b}
\end{align} 
\hrulefill 
\end{figure*}

Upon using \cite[Lemma 2]{soleymani2025framework} and the lower bounds in Lemma \ref{lem:1}, we can convert \eqref{opt} to the following non-FMP problem
\begin{subequations}\label{opt-w2}
    \begin{align}    
	\max\limits_{\{{\bf \Gamma}_k,\gamma_k\}_{\forall k}} & \sum_{k }\left(2\zeta_k^{(n)}\gamma_k-\zeta_k^{(n)^2}\left(P_s+\text{Tr}({\bf \Gamma}_k{\bf \Gamma}_k^H )\right) \right) \!\! \label{eq:opt} \\
	\label{eq:qos}\textrm{s.t.}\hspace{.5cm} & \tilde{r}_k\geq \bar{r}_k^{th}\\
    & \sum_{k=1}^K\text{Tr}\left( {\bf \Gamma}_k{\bf \Gamma}_k^H \right)\! \leq\! P 
    \\
    &\tilde{r}_k-\gamma_k^2\geq 0,
\end{align}
\end{subequations}
where we have $\zeta_k^{(n)}=\frac{\sqrt{\alpha_kr_k^{(n)}}}{P_s+\text{Tr}({\bf \Gamma}_k^{(n)}{\bf \Gamma}_k^{(n)^H} )}$, and $\gamma_k$ is an auxiliary variable for all $k$. 
 The problem in \eqref{opt-w2} is convex, solvable by standard toolboxes for solving convex problems such as CVX \cite{boyd2004convex}. The solution of \eqref{opt-w2} gives ${\bf \Gamma}_k^{(n+1)}$ for all $k$.

\subsection{RIS Optimization}

When the beamforming matrices are kept fixed to ${\bf \Gamma}_k^{(n+1)}$ for all $k$, \eqref{opt} is simplified as
    \begin{align}    \label{opt-t}
	\max\limits_{{\bf \Theta}^r, {\bf \Theta}^t} & \sum_{k } \frac{ \alpha_k{r}_k }{P_s+\text{Tr}({\bf \Gamma}_k^{(n+1)}{\bf \Gamma}_k^{(n+1)^H}) }  &
	\textrm{s.t.}\hspace{.5cm} &  \eqref{eq:qosi}, \eqref{eq:con2}. 
\end{align}
The optimization problem in \eqref{opt-t} is not an FMP problem, but still non-convex, since the rates are non-concave in ${\bf \Theta}^r$ and ${\bf \Theta}^t$. 
To address this issue, we utilize an approach based on majorization minimization (MM). That is, we derive lower bounds for $r_k$, which are concave in ${\bf \Theta}^r$ and ${\bf \Theta}^t$ as
\begin{multline}
    r_{k} \geq \hat{r}_{k}= a_{k}
+2\sum_{j}\mathfrak{R}\left\{\text{{Tr}}\left(
{\bf A}_{kj}{\bf \Gamma}_{j}^{(n+1)^H}
\mathbf{H}_{k}^H\right)\right\}
\\-
\text{{Tr}}\left(
{\bf B}_{k}\left(\sigma^2{\bf I}
+\sum_j\mathbf{H}_{k}{\bf \Gamma}_{j}^{(n+1)}{\bf \Gamma}_{j}^{(n+1)^H}\mathbf{H}_{k}^H\right)
\right),
\end{multline}
where the coefficients $a_{k}$, ${\bf A}_{kj}$, and ${\bf B}_{k}$ are, respectively, given by \eqref{eq-a}, \eqref{eq-Akj}, and \eqref{eq-b} on the top of the next page, except that  ${\bf \Gamma}_{k}^{(n)}$ is replaced by ${\bf \Gamma}_{k}^{(n+1)}$ for all $k$. In other words, the coefficients $a_{k}$, ${\bf A}_{kj}$, and ${\bf B}_{k}$ should be updated at each step based on the latest value of ${\bf \Theta}^r$, ${\bf \Theta}^t$, and ${\bf \Gamma}_{k}$ for all $k$.
Upon replacing $r_k$ with $\hat{r}_k$ in \eqref{opt-t}, we have
\begin{subequations}
    \label{opt-t2}
\begin{align}    
	\max\limits_{{\bf \Theta}^r, {\bf \Theta}^t}\, & \sum_{k } \frac{\alpha_k\hat{r}_k }{P_s+\text{Tr}({\bf \Gamma}_k^{(n+1)}{\bf \Gamma}_k^{(n+1)^H}) }  \\
	\textrm{s.t.}\hspace{.5cm} &  \eqref{eq:con2}, 
    \\& \hat{r}_k\geq\bar{r}_k^{th}, \,\forall k,
\end{align}
\end{subequations}
which is convex. Solving \eqref{opt-t2} gives ${\bf \Theta}^{r^{(n+1)}}$ and ${\bf \Theta}^{t^{(n+1)}}$.

\section{Numerical Results}
\label{sec:ni}

We employ Monte Carlo simulations to evaluate the energy-efficient scheme proposed in this work. In the simulation setup, we assume a line-of-sight (LOS) link for the channels ${\bf G}$ and ${\bf F}_k$ for all $k$, making their corresponding small-scale fading Ricean. In line with \cite[(55)]{pan2020multicell}, we assume a Ricean factor of three. Additionally, we assume that ${\bf D}_k$ is non-LOS, following the Rayleigh small-scale fading. Moreover, we model the large-scale fading of all channels according to the path-loss model in \cite[(59)]{soleymani2022improper}. The other propagation parameters, including the antenna gains, bandwidth, noise power density, path loss components, and the path loss at the reference distance of $1$ meter, are based on \cite{soleymani2022improper}. In the MS scheme, we consider $m_g=M/2$, implying that we divide the STAR-RIS elements into two groups with same number of elements.  Finally, we set $N_{BS}=N_u=2$, $K=6$, $l=256$ bits, and $\epsilon= 10^{-5}$, unless it is explicitly mentioned otherwise.

In the simulation scenario, half of the users are in the reflection space, while the other half are in the refraction space, as illustrated in Fig. \ref{fig:network-model}. Hence, a reflective RIS can aid only half of the users. In the figures, {\bf S-RIS}, {\bf S-RIS-MS},  {\bf RIS-Rand}, and {\bf RIS} refer to our energy-efficient algorithm conceived for STAR-RIS along with the ES scheme, STAR-RIS with the MS scheme, STAR-RIS with random coefficients, and reflective RIS, respectively. Finally, {\bf No-RIS} refers to the energy-efficient design for systems operating without RIS.

\begin{figure}
    \centering
    \includegraphics[width=.9\linewidth]{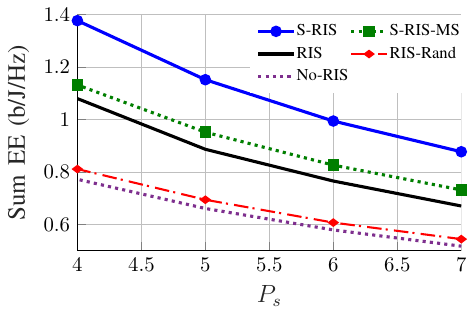}
    \caption{Sum EE versus $P_s$.}\vspace{-.5cm}
    \label{fig:sim1}
\end{figure}
Fig. \ref{fig:sim1} shows the average sum EE versus $P_s$, illustrating the superiority of STAR-RIS over reflective RIS in this simulation setup along with $M=60$. We observe that an RIS (either STAR or reflective) improves the EE if its elements are optimized according to the energy-efficient scheme proposed in this paper.  However, STAR-RIS has limited gains when its elements are chosen randomly. 

\begin{figure}
    \centering
    \includegraphics[width=.9\linewidth]{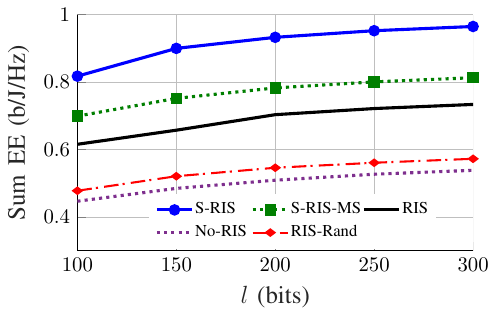}
    \caption{Sum EE versus  $l$.}\vspace{-.5cm}
    \label{fig:sim2}
\end{figure}
Fig. \ref{fig:sim2} shows the average sum EE versus the codeword length $l$ for $M=80$. When there is a demanding latency constraint, we have to utilize a shorter block length. Indeed, this figure demonstrates how energy-efficiently the system operates when the latency requirement is more stringent. In this example, STAR-RIS significantly improves the EE, and its benefits increase as $l$ reduces, meaning that STAR-RIS is even more beneficial in low-latency systems.

\begin{figure}
    \centering
    \includegraphics[width=.9\linewidth]{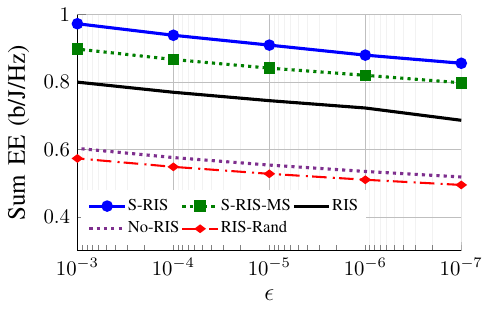}
    \caption{Sum EE versus $\epsilon$.}\vspace{-.5cm}
    \label{fig:sim3}
\end{figure}

Fig. \ref{fig:sim3} shows the average sum EE versus the maximum tolerable detecting error $\epsilon$ for $M=80$. The more reliable the system should operate, the less energy-efficient the network becomes. However, for all values of $\epsilon$, reflective and STAR RISs substantially improve EE. Moreover, STAR-RIS with the ES scheme outperforms the other algorithms. In this example, the gains provided by STAR-RIS are higher when $\epsilon$ decreases.

\section{Conclusions and Future Work}
\label{sec:conc}
We proposed an energy-efficient scheme for STAR-RIS-assisted MU-MIMO URLLC BCs, maximizing the sum EE of users. Our results illustrated that both the STAR-RIS and reflective RIS architectures can substantially increase the sum EE of the network. Moreover, STAR-RIS outperforms reflective RIS when the users are located in both reflection and refraction spaces. Indeed, STAR-RIS is more suitable when the RIS cannot be located in a position covering all the users in the reflection space. Additionally, STAR-RIS provides higher gains when there are more demanding latency and reliability constraints. 

\section*{Acknowledgment}\vspace{-.1cm}
I. Santamaria’s work was funded by MCIN/AEI/10.13039/ 501100011033, under Grant PID2022-137099NBC43 (MADDIE) and FEDER UE, and by European Union’s (EU’s) Horizon Europe project 6G-SENSES under Grant 101139282. E. Jorswieck’s work was supported in part by the Federal Ministry of Education and Research (BMBF, Germany) through the Program of Souver\"an. Digital. Vernetzt. joint Project 6G-RIC, under Grant 16KISK031, and by European Union's (EU's) Horizon Europe project 6G-SENSES under Grant 101139282. 
Robert Schober’s work was funded by the German Ministry for Education and Research (BMBF) under the program of ``Souver\"an. Digital. Vernetzt.'' joint project 6G-RIC (Project-ID 16KISK023). L. Hanzo would like to acknowledge the financial support of the Engineering and Physical Sciences Research Council (EPSRC) projects under grant EP/Y026721/1, EP/W032635/1, EP/Y037243/1 and EP/X04047X/1 as well as of the European Research Council's Advanced Fellow Grant QuantCom (Grant No. 789028).
\bibliographystyle{ieeetr}
\bibliography{ref.bib}

\begin{thebibliography}{10}

\bibitem{wang2023road}
C.-X. Wang {\em et~al.}, ``On the road to {6G}: Visions, requirements, key technologies and testbeds,'' {\em IEEE Commun. Surv. Tutor.}, vol.~25, no.~2, pp.~905--974, 2023.

\bibitem{gong2022holographic}
T.~Gong {\em et~al.}, ``Holographic {MIMO} communications: Theoretical foundations, enabling technologies, and future directions,'' {\em IEEE Commun. Surv. Tutor.}, vol.~26, no.~1, pp.~196--257, 2024.

\bibitem{vaezi2022cellular}
M.~Vaezi {\em et~al.}, ``Cellular, wide-area, and non-terrestrial {IoT}: A survey on {5G} advances and the road towards {{{6G}}},'' {\em IEEE Commun. Surv. Tutor.}, vol.~24, no.~2, pp.~1117--1174, 2022.

\bibitem{wu2021intelligent}
Q.~Wu {\em et~al.}, ``Intelligent reflecting surface aided wireless communications: A tutorial,'' {\em IEEE Trans. Commun.}, vol.~69, no.~5, pp.~3313--3351, 2021.

\bibitem{di2020smart}
M.~Di~Renzo {\em et~al.}, ``Smart radio environments empowered by reconfigurable intelligent surfaces: How it works, state of research, and the road ahead,'' {\em IEEE J. Sel. Areas Commun.}, vol.~38, no.~11, pp.~2450--2525, 2020.

\bibitem{lyu2023energy}
W.~Lyu, Y.~Xiu, S.~Yang, C.~Yuen, and Z.~Zhang, ``Energy-efficient cell-free network assisted by hybrid {RIS}s,'' {\em IEEE Wireless Commun. Lett.}, vol.~12, no.~4, pp.~718--722, 2023.

\bibitem{soleymani2024energy}
M.~Soleymani, I.~Santamaria, E.~Jorswieck, M.~Di~Renzo, and J.~Guti{\'e}rrez, ``Energy efficiency comparison of {RIS} architectures in {MISO} broadcast channels,'' in {\em IEEE 25th International Workshop on Signal Processing Advances in Wireless Communications (SPAWC)}, pp.~701--705, IEEE, 2024.

\bibitem{soleymani2022rate}
M.~Soleymani, I.~Santamaria, and E.~Jorswieck, ``Rate splitting in {{MIMO}} {{RIS}}-assisted systems with hardware impairments and improper signaling,'' {\em IEEE Trans. Veh. Technol.}, vol.~72, no.~4, pp.~4580--4597, April 2023.

\bibitem{niu2023active}
H.~Niu {\em et~al.}, ``Active {RIS} assisted rate-splitting multiple access network: Spectral and energy efficiency tradeoff,'' {\em IEEE J. Sel. Areas Commun.}, vol.~41, no.~5, pp.~1452--1467, 2023.

\bibitem{soleymani2022noma}
M.~Soleymani, I.~Santamaria, E.~Jorswieck, and S.~Rezvani, ``{NOMA}-based improper signaling for multicell {MISO} {{RIS}}-assisted broadcast channels,'' {\em IEEE Trans. Signal Process.}, vol.~71, pp.~963--978, 2023.

\bibitem{fotock2023energy}
R.~K. Fotock, A.~Zappone, and M.~Di~Renzo, ``Energy efficiency optimization in {RIS}-aided wireless networks: Active versus nearly-passive {RIS} with global reflection constraints,'' {\em IEEE Trans. Commun.}, vol.~72, no.~1, pp.~257--272, 2024.

\bibitem{polyanskiy2010channel}
Y.~Polyanskiy, H.~V. Poor, and S.~Verd{\'u}, ``Channel coding rate in the finite blocklength regime,'' {\em IEEE Trans. Inf. Theory}, vol.~56, no.~5, pp.~2307--2359, 2010.

\bibitem{erseghe2016coding}
T.~Erseghe, ``Coding in the finite-blocklength regime: Bounds based on laplace integrals and their asymptotic approximations,'' {\em IEEE Trans. Inf. Theory}, vol.~62, no.~12, pp.~6854--6883, 2016.

\bibitem{soleymani2024rate2}
M.~Soleymani, A.~Zappone, E.~Jorswieck, M.~Di~Renzo, and I.~Santamaria, ``Rate region of {RIS}-aided {URLLC} broadcast channels: Diagonal versus beyond diagonal globally passive {RIS},'' {\em IEEE Wireless Commun. Let.}, vol.~14, no.~2, pp.~320--324, 2025.

\bibitem{li2021aerial}
Y.~Li, C.~Yin, T.~Do-Duy, A.~Masaracchia, and T.~Q. Duong, ``Aerial reconfigurable intelligent surface-enabled {URLLC} {UAV} systems,'' {\em IEEE Access}, vol.~9, pp.~140248--140257, 2021.

\bibitem{vu2022intelligent}
T.-H. Vu, T.-V. Nguyen, D.~B. da~Costa, and S.~Kim, ``Intelligent reflecting surface-aided short-packet non-orthogonal multiple access systems,'' {\em IEEE Trans. Veh. Technol.}, vol.~71, no.~4, pp.~4500--4505, 2022.

\bibitem{xie2021user}
H.~Xie, J.~Xu, Y.-F. Liu, L.~Liu, and D.~W.~K. Ng, ``User grouping and reflective beamforming for {IRS}-aided {URLLC},'' {\em IEEE Wireless Commun. Lett.}, vol.~10, no.~11, pp.~2533--2537, 2021.

\bibitem{almekhlafi2021joint}
M.~Almekhlafi, M.~A. Arfaoui, M.~Elhattab, C.~Assi, and A.~Ghrayeb, ``Joint resource allocation and phase shift optimization for {{RIS}}-aided {eMBB/URLLC} traffic multiplexing,'' {\em IEEE Trans. Commun.}, vol.~70, no.~2, pp.~1304--1319, 2022.

\bibitem{soleymani2023spectral}
M.~Soleymani, I.~Santamaria, and E.~Jorswieck, ``Spectral and energy efficiency maximization of {MISO} {STAR-{RIS}}-assisted {URLLC} systems,'' {\em IEEE Access}, vol.~11, pp.~70833--70852, 2023.

\bibitem{pala2023spectral}
S.~Pala, M.~Katwe, K.~Singh, B.~Clerckx, and C.-P. Li, ``Spectral-efficient {RIS}-aided {RSMA} {URLLC}: Toward mobile broadband reliable low latency communication ({mBRLLC}) system,'' {\em IEEE Trans. Wireless Commun.}, vol.~23, no.~4, pp.~3507--3524, 2024.

\bibitem{soleymani2023optimization}
M.~Soleymani, I.~Santamaria, E.~Jorswieck, and B.~Clerckx, ``Optimization of rate-splitting multiple access in beyond diagonal {RIS}-assisted {URLLC} systems,'' {\em IEEE Trans. Wireless Commun.}, vol.~23, no.~5, pp.~5063--5078, 2024.

\bibitem{katwe2024rsma}
M.~V. Katwe, R.~Deshpande, K.~Singh, M.-L. Ku, and B.~Clerckx, ``{RSMA}-enabled aerial {RIS}-aided {MU}-{MIMO} system for improved spectral-efficient {URLLC},'' {\em IEEE Trans. Veh. Technol.}, vol.~74, no.~2, pp.~3110--3127, 2025.

\bibitem{soleymani2024optimization}
M.~Soleymani, I.~Santamaria, E.~Jorswieck, R.~Schober, and L.~Hanzo, ``Optimization of the downlink spectral-and energy-efficiency of {RIS}-aided multi-user {URLLC} {MIMO} systems,'' {\em IEEE Trans. Commun.}, vol.~73, no.~5, pp.~3497--3513, 2025.

\bibitem{soleymani2024rate}
M.~Soleymani, I.~Santamaria, E.~Jorswieck, M.~Di~Renzo, R.~Schober, and L.~Hanzo, ``Rate splitting multiple access for {RIS}-aided {URLLC} {MIMO} broadcast channels,'' {\em Submitted to IEEE Trans. Wireless Commun.}, 2024.

\bibitem{zhang2022intelligent}
H.~Zhang and B.~Di, ``Intelligent omni-surfaces: Simultaneous refraction and reflection for full-dimensional wireless communications,'' {\em IEEE Commun. Surv. Tutor.}, vol.~24, no.~4, pp.~1997--2028, 2022.

\bibitem{liu2021star}
Y.~Liu, X.~Mu, J.~Xu, R.~Schober, Y.~Hao, H.~V. Poor, and L.~Hanzo, ``{STAR}: Simultaneous transmission and reflection for 360 coverage by intelligent surfaces,'' {\em IEEE Wireless Commun.}, vol.~28, no.~6, pp.~102--109, 2021.

\bibitem{jorswieck2025urllc}
E.~Jorswieck, M.~Soleymani, I.~Santamaria, and J.~Guti{\'e}rrez, ``{URLLC} networks enabled by {STAR}-{RIS}, rate splitting, and multiple antennas,'' in {\em Proc. IEEE Int. Conf. Mobile Miniaturized Terahertz Syst. (ICMMTS)}, pp.~1--5, 2025.

\bibitem{mu2021simultaneously}
X.~Mu, Y.~Liu, L.~Guo, J.~Lin, and R.~Schober, ``Simultaneously transmitting and reflecting ({STAR}) {{RIS}} aided wireless communications,'' {\em IEEE Trans. Wireless Commun.}, vol.~21, no.~5, pp.~3083--3098, 2022.

\bibitem{liu2022star}
Y.~Liu, X.~Mu, J.~Xu, R.~Schober, Y.~Hao, H.~V. Poor, and L.~Hanzo, ``{STAR}: Simultaneous transmission and reflection for 360° coverage by intelligent surfaces,'' {\em IEEE Wireless Commun.}, vol.~28, no.~6, pp.~102--109, 2021.

\bibitem{soleymani2023noma2}
M.~Soleymani, I.~Santamaria, and E.~Jorswieck, ``{NOMA}-based improper signaling for {MIMO} {STAR}-{RIS}-assisted broadcast channels with hardware impairments,'' in {\em IEEE Global Communications Conference (GLOBECOM)}, pp.~673--678, IEEE, 2023.

\bibitem{soleymani2023energy}
M.~Soleymani, I.~Santamaria, and E.~Jorswieck, ``Energy-efficient rate splitting for {{MIMO}} {STAR}-{{RIS}}-assisted broadcast channels with {I/Q} imbalance,'' {\em Proc. IEEE Eu. Signal Process. Conf. (EUSIPCO)}, 2023.

\bibitem{soleymani2022improper}
M.~Soleymani, I.~Santamaria, and P.~J. Schreier, ``Improper signaling for multicell {{MIMO}} {{RIS}}-assisted broadcast channels with {I/Q} imbalance,'' {\em IEEE Trans. Green Commun. Netw.}, vol.~6, no.~2, pp.~723--738, 2022.

\bibitem{soleymani2025framework}
M.~Soleymani, E.~Jorswieck, R.~Schober, and L.~Hanzo, ``A framework for fractional matrix programming problems with applications in {FBL} {MU}-{MIMO},'' {\em Submitted to IEEE Trans. Wireless Commun.}, 2025.

\bibitem{boyd2004convex}
S.~Boyd and L.~Vandenberghe, {\em Convex {O}ptimization}.
\newblock Cambridge University Press, 2004.

\bibitem{pan2020multicell}
C.~Pan, H.~Ren, K.~Wang, W.~Xu, M.~Elkashlan, A.~Nallanathan, and L.~Hanzo, ``Multicell {{MIMO}} communications relying on intelligent reflecting surfaces,'' {\em IEEE Trans. Wireless Commun.}, vol.~19, no.~8, pp.~5218--5233, 2020.

\end{thebibliography}

\end{document}